# Amido Rhenium Trioxides: Cases of Hindered Agostic C-H··M Interactions?

Paul Benndorf,[a] Michael T. Gamer,[a] Peter W. Roesky,*[a] Georg Eickerling*[b] and Wolfgang Scherer*[b]

___________________________

[a]  Prof. Dr. P. W. Roesky, Dipl.-Chem. P. Benndorf, Dr. M. Gamer

   Institut für Anorganische Chemie

   Universität Karlsruhe (TH)

   Engesserstr. 15, 76131 Karlsruhe (Deutschland)

   Fax: (+49)721-608-4854

   E-mail: roesky@chemie.uni-karlsruhe.de

[b]  Prof. Dr. W. Scherer, Dr. G. Eickerling

   Institut für Physik

   Lehrstuhl für Chemische Physik und Materialwissenschaften

   Universität Augsburg, 86135 Augsburg (Deutschland)

   Fax: (+49)821-598-3227

   E-mail: wolfgang.scherer@physik.uni-augsburg.de

   Supporting information for this article is available from the author

**Abstract.**

The amido rhenium trioxides of composition ($i$Pr$_2$N)ReO$_3$, ($i$PrCyN)ReO$_3$ and (Cy$_2$N)ReO$_3$ (Cy = cyclohexyl) were synthesized in a one pot reaction starting from Re$_2$O$_7$, Me$_3$SiCl and the corresponding amines ($i$Pr)$_2$NH, ($i$Pr)(Cy)NH, and (Cy)$_2$NH, respectively. Even two of the three compounds are high viscous oils or waxes at room temperature single crystals could be obtained at low temperatures and the solid state structures were established by single crystal X-ray diffraction. In the solid state the amido ligands of all three complexes are asymmetrically coordinated to the ReO$_3$ core allowing for one short Re$\cdots$H-C contact in each case which might indicate the presence of β-agostic interaction. However, analysis of the charge density distribution provided us clear-cut criteria that β-agostic interactions are suppressed by the *trans*-influence of the oxo-groups. Comparison with structurally related tetra-coordinated $d^0$ titanium amido complexes highlighted a systematic concept how the extent of β-agostic interactions in these complexes can be controlled by reducing the *trans*-influence of the co-ligands. We therefore suggest to employ the expression "hindered agostic interactions" in cases where covalent M$\cdots$H-C are in principle supported by vacant coordination sites at Lewis-acidic metal centers but are actually hindered for electronic reasons. We further point out that electronic hindrance originating from the *trans*-influence might be overruled by changing the σ/π-donor characteristics of ligands. Hence, tetra-coordinated amido or alkyl $d^0$ complexes seemingly represent a class of compounds where C-H bond activation can be controlled in the electronic ground state geometries in a systematic way by designing the electronic characteristics of the coordinating co-ligands.

**Introduction**

High-oxidation-state transition metal oxides are attractive synthetic reagents and catalysts. The compounds can be used for highly controlled synthetic transformations to occur intramolecularly within the metal coordination sphere.[1] In rhenium(VII) chemistry a number of complexes of the general formula [LReO$_3$] (L = inorganic or organic substituent) have been investigated.[2, 3] The most prominent species in this series is methylrhenium trioxide CH$_3$ReO$_3$,[4] which has been used as catalyst in oxidation reaction as well as in olefin metathesis[5, 6, 7, 8, 9] and more recently as chimie douce precursor molecule for the synthesis of inorganic Re(IV) and Re(VI) oxides[10] or conducting organometallic oxides.[11] In contrast to the well established organometallic chemistry of rhenium(VII)oxides the corresponding amido complexes containing the [ReO$_3$]$^+$ core are extremely rare. Beside phosphiminato derivatives, (O$_3$ReN=PR$_3$),[12] the benzamidinate compound [{(CH$_3$)$_3$SiNC(Ph)NSi(CH$_3$)$_3$}ReO$_3$],[13] and some $\eta^2$-pyrazolate complexes of the type [($\eta^2$-pz)ReO$_3$(THF)$_n$] (pz = pyrazolate)[14] only two amido rhenium trioxides of the composition (R$_2$N)ReO$_3$ (R = $i$Pr, Mesityl (Mes)) were reported by G. Wilkinson and co-workers about two decades ago.[15, 16] The first one, ($i$Pr$_2$N)ReO$_3$ was reported being a thermally unstable compound, which decomposes both in the solid state and in solution ($t_{1/2}$ *ca.* 1 d, 20 °C).[15] The second and only structural characterized amido rhenium trioxide, (Mes$_2$N)ReO$_3$, was obtained in a peculiarly way upon reaction of Mes$_2$ReO$_2$ with NO.[16] In contrast, the less bulky compound Me$_2$NReO$_3$ could not be isolated.[15] Based on these result we desired to synthesize some amido rhenium trioxides, (R$_2$N)ReO$_3$ which are sufficiently stabilized to allow us the exploration of their solid state geometries and electronic structures. We were especially interested to investigate whether the presence of β(C-H) groups in the alkyl backbones already induces β-agostic interactions between the alkyl backbone and the Lewis-acidic Re$^{VII}$ metal centers. Comparison with related $d^0$ amido and alkyl metal complexes should then provide further insight which control parameter trigger C-H activation processes in $d^0$ transition metal complexes at the early stage of the reaction coordinate of the β-hydride elimination channel.

## Results and Discussion

**Preparation and Characterization of Amido Rhenium Oxides**

The synthesis of the amido rhenium trioxides was performed in a one pot reaction. In the first step Cl-ReO$_3$ was generated *in situ* according to a modified literature procedure starting from Re$_2$O$_7$ and Me$_3$SiCl (Scheme 1).[17] The *in situ* generated Cl-ReO$_3$ was further reacted at -78 °C with the amines (*i*Pr)$_2$NH, (*i*Pr)(Cy)NH, and (Cy)$_2$NH (Cy = cyclohexyl), respectively, to give the corresponding amido rhenium trioxides (RR´N)ReO$_3$ (R, R´= *i*Pr (**1**); R = *i*Pr, R´= Cy (**2**); R, R´= Cy (**3**)) in reasonable yields (Scheme 1). Compounds **1-3** have been characterized by standard analytical / spectroscopic techniques.

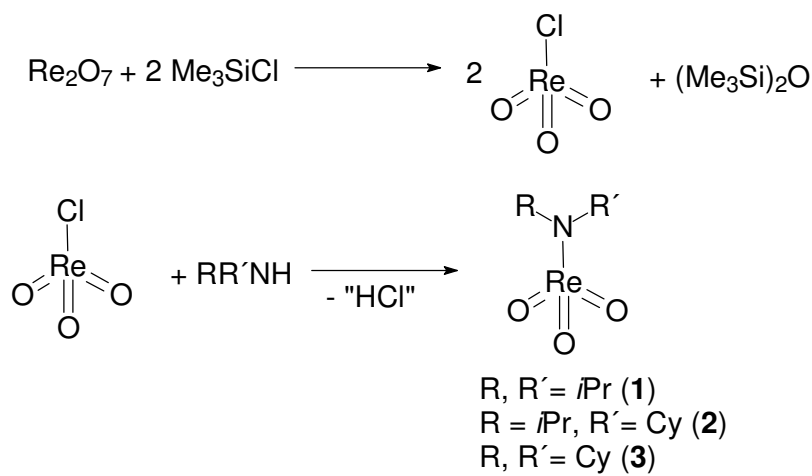

R, R´= *i*Pr (**1**)
R = *i*Pr, R´= Cy (**2**)
R, R´= Cy (**3**)

**Scheme 1**

The IR spectrum of compound **1** shows two intense Re=O stretching modes at 977 cm$^{-1}$ and 934 cm$^{-1}$. A characteristic septet (δ 3.40 ppm) and the expected doublet (δ 0.77 ppm) are observed for the isopropyl groups in the $^1$H NMR spectrum of compound **1**. Even though compound **1** is a yellow oil at room temperature, single crystals could be obtained from cold pentane solutions. The solid state structure of complex **1** was established by single crystal X-ray diffraction (Figure 1). Most interesting is the asymmetric coordination of the diisopropyl ligand onto the ReO$_3$ core. One isopropyl group is significantly tilted towards the Re atom resulting in asymmetric C-N-Re angles of C1-N-Re 125.6(5)° and C4-N-Re 114.7(5)°. In contrast the ReO$_3$ core shows the expected bonding parameters. Thus, the Re-O bond distances are in the typical range of Re(VII)=O double bonds (1.694(8) Å - 1.710(6) Å)[18] and the Re-N bond length of compound **1** (1.893(6) Å) is similar to the one observed in (Mes$_2$N)ReO$_3$ (1.909(10) Å) [16] but clearly shorter than the σ(Re-C) bond in MTO (2.063(2)Å).[11c] Hence, the π-donor capability of the diisopropyl ligand is dominant in **1** which is also reflected in the virtual planarity of the ReNC$_2$ fragment.

**Figure 1 here**

As a result of the higher steric demand of the *N*-cyclohexylisopropylamido relative to the diisopropylamido ligand compound **2** is slightly more stable than compound **1**. The $^1$H NMR spectrum shows the expected set of broad signals and the characteristic triplet of triplets of the N-C*H* group of the cyclohexyl ring. Moreover, the characteristic septet (δ 3.20 ppm) and the expected doublet (δ 0.82 ppm) are observed for the isopropyl group. In the IR spectrum of compound **2** two intense Re=O vibrations bands are seen at 976 cm$^{-1}$ and 931 cm$^{-1}$.

Compound **2** is a wax at room temperature but crystallizes from cold pentane solutions in the orthorhombic space group *Pbca* having two molecules in the asymmetric unit. These two molecules differ by the arrangement of the amido ligand (Figure 2). As observed in compound **1** the amido ligand is asymmetrically tilted towards the ReO$_3$ core. Whereas, in compound **1** both R groups are equivalent, in compound **2** an isopropyl and a cyclohexyl group are attached to the amido nitrogen atom. Thus, in the solid state structure the isopropyl group is bent to Re1 (C1-N1-Re1 113.4(3)°) in the first isomer whereas the cyclohexyl group is close to Re2 (C13-N2-Re2 116.3(2)°) in the second isomer. In both isomers the cyclohexyl group adopts a chair type conformation. The distorted tetrahedral coordination polyhedra of the rhenium atoms are formed by the three oxygen atoms and the nitrogen atom. The bond distances and angles of both isomers within the NReO$_3$ core are in a similar range. Thus, the Re-O bond distances are Re1-O1 1.696(3) Å, Re1-O2 1.694(3) Å, Re1-O3 1.704(4) Å; Re2-O4 1.698(3) Å, Re2-O5 1.699(3) Å, Re2-O6 1.689(4) Å.

**Figure 2 here**

Compound **3**, which was obtained by reaction of Cl-ReO$_3$ and (Cy)$_2$NH in THF, is a solid at room temperature. Compound **3** was characterized by IR, $^1$H, and $^{13}$C{$^1$H} NMR spectroscopy. In the $^1$H NMR spectrum broad signals in the range of 1.40 ppm and 0.77 ppm for the cyclohexyl ring but also a characteristic triplet of triplets of the N-C*H* group is seen at δ 3.31 ppm. The expected four resonances are observed in the

$^{13}$C{$^1$H} NMR spectrum. As observed for compounds **1** and **2** also IR spectrum of compound **3** displays both expected intense Re=O stretching modes at 951 cm$^{-1}$ and 921 cm$^{-1}$. By using mass spectrometry a molecular peak can be detected.

Single crystals of complex **3** could be obtained from a pentane solution at -22 °C. The solid-state structure of **3** was investigated by single crystal X-ray diffraction. Compound **3** crystallizes in the orthorhombic space group *Pna*2$_1$ as racemic twin. As observed for compounds **1** and **2** the amido ligand again is asymmetrically attached to the metal center showing two different bonding angles of C1-N-Re 114.0(2)° and C7-N-Re 127.7(12)°, which differ by more then 13°. The Re-N bond length of compound **3** (1.876(2) Å) is similar to the one observed in 1 (1.893(6) Å) and 2 (1.876(3) Å). The N-ReO$_3$ core forms as expected a distorted tetrahedron.

**Figure 3 here**

Compound **3** is as a result of the larger steric demand more stable than compounds **1** and **2**. Thus, compound **3** is stable in air at ambient conditions for a couple of days. On the other hand **3** slowly decomposes in solution, which is an explanation for the relative low yields obtained for complexes **1-3**. The amines applied for the synthesis of the amido rhenium trioxides are somewhat related. Attempts to use other amines failed. By using more sterically demanding amines such as 2,2,6,6-tetramethylpiperidin, di*tert*-butylamine, and *N*-isopropyl-*tert*-butylamine no reaction was observed. In contrast, reactions of Cl-ReO$_3$ with sterically less demanding amines such as diphenylamine and iminostilbene resulted in rapid decomposition in solution even at low temperatures. The yellow solutions (-78 °C) turned black upon warming.

**Electronic Structure Analysis**

In this section we will focus on the electronic situation of complex **1** which is also characteristic for the related amide complexes **2** and **3**. The diisopropylamido (DA) ligand (*i*Pr$_2$N$^-$) employed in complex **1** has an outstanding position in organometallic chemistry due to its sterical demand and its π-donor capabilities. Despite its susceptibility to β-hydride elimination caused by the presence of the methine hydrogen atom it has been successfully employed to isolate reactive homoleptic transition metal isoproylamides such as [(*i*Pr$_2$N)$M^{II}$(μ-(*i*Pr$_2$N))]$_2$ ($M$ = Cr ($d^4$), Mn($d^5$)).[19] Hence, the presence of β-hydrogen atoms in the amide ligand does not seem to be the only prerequisite for the thermal decomposition pathway leading to the elimination of the imine *i*Pr-N=C(Me)$_2$ and the corresponding metal hydride species. Indeed, the propagation of the classical β-hydride elimination process should be initiated by a preceding C-H activation step of a methine-hydrogen bond of the DA ligand leading to a covalent agostic M⋯H-C interaction. However, beside geometrical factors it is the electronic situation at the metal atoms which controls the C-H activation step and the subsequent β-hydrogen elimination pathway.[20] In this bonding scenario the elimination process can be considered as formal oxidative addition of a σ(C-H) bond[21] of the DA ligand at the metal center under formation of a metal-hydride bond and simultaneous release of the *i*Pr-N=C(Me)$_2$ imine. However, such reaction pathway involving an oxidative addition step is usually suppressed in case of $d^0$ type compounds such as **1** since the $d^0$ metal center cannot formally back-donate electrons to the σ*(C-H) bond. Hence, the alternative isolation of C-H activated $d^0$ intermediates displaying pronounced agostic Re⋯H-C interactions might be favored relative to the β-elimination products.[22]

At the first glance, the observation of a distorted DA ligand displaying an asymmetric pair of different ∠ReNC valence angles (114.7(5)° and 125.6(5)°) in the planar RNC$_2$ moiety hints for the presence of Re···H-C interaction in **1** (Figure 4). As a result of the ligand tilt the methine hydrogen atom of the iso-propyl group at the near side of the metal establishes a remarkably short Re···H-C contact of 2.66 [2.659] Å (values based on DFT calculations are specified in square brackets in the following). A similar scenario is observed for the structurally related Ti($d^0$) amide (DA)TiCpCl$_2$ **4** (Figure 4) where the asymmetry in the ∠TiNC angles of the planar TiNC$_2$ fragment is even more pronounced: ∠TiNC = 101.4(2)° and 146.1(2)°.[23] Note, that **4** and its Zr-analogue **4'** represent benchmark amido transition metal complexes characterized by the smallest ∠MNC angles and shortest M···H contacts which are reported in the CSD database[24] so far (Figure 5). Accordingly, the methine H atom at the near side of the titanium atom shows an even shorter M···H-C contact of 2.25 [2.323] Å relative to **1** which is close to the distance range of approximately 2.05-2.30 Å displayed by typical β-agostic $d^0$ complexes (Figure 5). Despite the fact that the Re···H$_β$ contact of 2.66 Å in **1** is somewhat larger than in **4** it is still significantly shorter than in the experimental gas phase models of EtTiCl$_3$ (Ti···H$_β$ = 3.24(2) Å)[25] and EtReO$_3$ (Re···H$_β$ = 3.16(5) Å) representing benchmarks for non-agostic $d^0$-configurated transition metal ethyl complexes. Hence, the unusual structure of **1** and the evidence of tilted methyl groups in the related $d^0$ compound trimethyldioxorhenium, (CH$_3$)$_3$ReO$_2$[26] suggest that delocalization giving the appearance of Re···H-C interactions is by no means impossible in these rhenium(VII) species with a valence electron count at, or close to, 18.[27]

**Figure 4 here**

Isolobal replacement of the demanding cyclopentadienyl moiety in **4** by a chloro ligand leads to our theoretical model complex (DA)TiCl$_3$ **5** which can be more directly compared with the rhenium complex **1**. **5** displays a calculated M⋯H distances of 2.442 Å which is clearly longer relative to that in **4** but still shorter than in **1**. This finding already indicates that subtle differences in the sterical demand and/or the electronic nature of the co-ligands might act as control parameter to manipulate the nature and extent of agostic M⋯H-C interactions in our amido benchmark systems **1**-**5**. At this stage, however, we should recall the electronic, spectroscopic and geometrical characteristics of agostic interactions in $d^0$ type complexes and realize that none of the five complexes **1**-**5** (and possibly all amido benchmark systems displayed in Figure 5) should be classified as truly *covalent* agostic species. Indeed, Figure 5 reveals that the geometrical characteristics of β-agostic $d^0$ transition metal alkyls (all display acute MCC valence angles) clearly deviate from those of the respective β-agostic amido congeners. This is an important fact which has– to the best of our knowledge – not been addressed so far in literature.

**Figure 5 here**

As pointed out above, covalent agostic interactions are characterized by a significant activation of the metal coordinating C-H group as a result of donation of charge density from the C-H bond to the Lewis-acidic metal center. A reduction in frequency for $\nu$(CH) is therefore the most direct observable of these changes, and the identification of a feature in the region 2800-2000 cm$^{-1}$ has in many instances been interpreted as diagnostic evidence of a M⋯H-C interaction.[20a,b,d,g] However, no low

ν(C-H) stretching frequencies (< 2800 cm$^{-1}$) signaling elongated C-H bonds have been reported so far for the tetrahedral Re($d^0$) species **1**-**3** and their structurally related Ti($d^0$) congener **4** whereas archetypal β-agostic complexes like [EtTiCl$_3$(dmpe)] show a significantly perturbed β-C-H agostic bond, with an isolated $ν^{is}$(CH) stretching frequency of 2585 cm$^{-1}$.[28] The lack of any significant C-H activation in the DA ligand in both benchmarks is supported by DFT calculations (Figure 4). Accordingly, the methine C-H bonds at the near side of the metal in **1** and **4** are only insignificantly elongated by 0.004 Å in comparison with the non-coordinated methine C-H group. Furthermore, the X-ray structural models as well as the DFT models of **1** and **4** do not indicate any significant shortening of N-C(methine) bond at the near side of the metal. The calculated electronic ground state geometries of **1** and **4** suggest that both systems are at the very early stage of the reaction coordinate of a β-hydride elimination reaction leading to a metal-hydride bond and the release of free *i*Pr-N=C(Me)$_2$.

**Figure 6 here**

As a result of our analysis true agostic interactions seem to be hindered in the $d^0$ Re/Ti amido systems **1**-**5** despite the fact that all classical prerequisites in favour of significant M←H-C donation are fulfilled: (*i*) -Lewis acidic character, (*ii*) low coordination numbers and (*iii*) presence of acceptor orbitals of suitable symmetry at the metal center.[20a,b] However, as outlined in a recent review, it is the microscopic local Lewis-acidity at the metal center which controls the extent of C-H activation in transition metal complexes.[20e] Indeed, topological analyses of the negative Laplacian of the charge density $L(\mathbf{r}) = -\nabla^2\rho(\mathbf{r})$, of **1** reveal the formation of local charge concentrations induced by the *trans*-influence of oxo-ligands at the metal site opposing the oxo groups but facing the N-C(methine) bond (Figure 6a). Hence, the local Lewis-acidity of the Re$^{VII}$ center is *locally* significantly reduced in the valence shell[29,30] by so-called ligand-induced charge concentrations (LICCs).[31] Hence, the Re←H-C interaction can be classified as an *electronically-hindered agostic interactions* at the early stage of the reaction coordinate of the β-hydride elimination process.

Accordingly, the extent of such a hindered agostic interaction might be controlled via manipulation of the magnitude of the *trans* influence of the opposing ligand. As exemplified by our model **5** - displaying a rather weak σ/π donor chloro ligand in *trans* position to the agostic C-H group - a smaller MNC angle of 107.0° (at the near side of the metal) and a shorter M⋯H distance of 2.442 Å relative to **1** (113.4° / 2.659 Å) and (DA)MnO$_3$ **6** (115.1° / 2.605 Å) is computed. Isolobal replacement of the chloro ligand in **5** by the Cp ring, lacking any significant *trans* influence, further lowers the MNC angle to 101.1° and shortens the M⋯H bond distance to 2.323 Å as calculated for compound **4**. Indeed, Figure 6b shows that the agostic C-H bond of **4** faces a zone of local charge depletion or locally increased Lewis-acidity at the metal center. Hence, the extent of the agostic interaction is largest in **4**, significantly reduced in **5** and hardly present in **1** and **6**.

We finally point out that also the amido ligand itself seems to hinder C(methine)-H activation. To the best of our knowledge, even in the case of coordinatively unsaturated and highly Lewis-acidic lanthanide DA complexes no acute ∠LnNC valence angles (Ln = lanthanide) or significantly reduced ν(C-H) stretching frequencies (<2800 cm$^{-1}$) have been described in the literature so far. Obviously, the pronounced π-bonding characteristics of the coordinating nitrogen atom hinders the amido ligand to form ∠LnNC valence angles smaller than 90°. This might also explain why the DA ligand is typically sufficiently stabilized towards β-H elimination processes and can be successfully employed to stabilize highly Lewis-acidic complexes displaying even $d$-electron counts ≥2. As a result, replacement of the ($i$Pr$_2$N$^-$) ligand in **4** by its isoelectronic ($i$Pr$_2$CH$^-$) alkyl congener leads to the new model complex ($i$Pr$_2$CH)TiCpCl$_2$ **7** (Figure 7) displaying a pronounced β-agostic interaction: ∠TiCC = 94.8°; Ti$\cdots$H$_\beta$ = 2.358 Å and C-H$_\beta$ = 1.111 Å (vs. 1.106 Å in **4**). Further elimination of the *trans* positioned Cp ligand results in the cationic model system ($i$Pr$_2$CH)TiCl$_2^+$ **8** (Figure 7) which shows one of the largest C-H bond activation ($r$(C-H) = 1.200 Å ) reported so far for a $d^0$-type complex. This result stresses again that the extent of β-agostic interactions in $d^0$ configured transition metal complexes can be *systematically controlled* by the local Lewis-acidity of the transition metal center and the efficient delocalization of the bonding M-C electron pair over the alkyl backbone. The latter process (negative hyperconjugation) which is pronounced in the distorted alkyl-backbones of β-agostic metal complexes appears hindered in the case of amido ligands.[20d,e,32]

**Figure 7 here**

**Summary.**

In summary, we have synthesized and structurally characterized the amido rhenium(VII) trioxides of composition ($i$Pr$_2$N)ReO$_3$ , ($i$PrCyN)ReO$_3$ and (Cy$_2$N)ReO$_3$ (**1-3**). All three compounds can be isolated in form of single crystals from cold pentane solutions. However, like EtReO$_3$ also these new amido complexes can not compete with the unprecedented chemical stability of the Re(VII) oxide benchmark $CH_3ReO_3$. Indeed, all amido rhenium trioxides decompose in solution more or less rapidly, possibly due to the presence of β-hydrogen atoms and remarkably short Re⋯H-C contacts (*aprox*. 2.6 Å). The presence of such β-agostic interactions appears indicated by the pronounced tilt of the amido ligand featuring pairs of small and large ∠ReNC angles (113.4(3)°-116.3(2)° vs. 125.6(3)°-127.7(3)°) in **1**-**3**. However, analysis of the charge density distribution provided us clear-cut criteria that β-agostic interactions in **1**-**3** are suppressed by the *trans*-influence of the oxo-groups. Comparison of the electronic situation in **1**-**3** with that in structurally related four-coordinated $d^0$ amido complexes revealed a systematic concept how the extent of β-agostic interactions in $d^0$ amido complexes can be controlled by reducing the *trans*-influence of the co-ligands. We therefore suggest to employ the expression "hindered agostic interactions" in cases like **1**-**3** where covalent M⋯H-C are in principle supported by vacant coordination sites at Lewis-acidic metal centers but are actually hindered for electronic reasons. Direct comparison of the β-agostic amido and alkyl complexes further revealed that the extent of C-H activation is significantly more pronounced in the latter ones due to the hindrance of negative hyperconjugative delocalization of the M-N bonding electrons over the ligand backbone. Hence, β-elimination processes seem to play a minor role in case of amido ligands. This again appears to be the true reason for the outstanding position of the diisopropyl amido ligand in organometallic chemistry besides its sterical demand and π-donor capabilities.

**Experimental Section.**

*General considerations:* All manipulations of air-sensitive materials were performed with the rigorous exclusion of oxygen and moisture in flame-dried Schlenk-type glassware on a dual manifold Schlenk line or in an argon-filled MBraun glove box. Ether solvents (THF and ethyl ether) were predried over Na wire and distilled under nitrogen from K (THF) as well as benzophenone ketyl prior to use. Hydrocarbon solvents (toluene and *n*-pentane) were distilled under nitrogen from $LiAlH_4$. Deuterated solvents were obtained from Chemotrade Chemiehandelsgesellschaft mbH (all ≥ 99 atom % D) and were degassed, dried, and stored *in vacuo* over Na/K alloy in resealable flasks. NMR spectra were recorded on JNM-LA 400 FT-NMR spectrometer. Chemical shifts are referenced to internal solvent resonances and are reported relative to tetramethylsilane. Mass spectra were recorded with a Varian MAT 771 spectrometer using the electronic impact technique (EI) with an ionization energy of 80 eV. IR spectra were obtained with a Shimadzu FTIR 8400S spectrometer. $Re_2O_7$ was purchased from Alfa Aesar. The amines were purchased from Acros Organics and were used without further purification.

**General procedure for the preparation of 1-3.**

$Me_3SiCl$ (0.13 mL, 1.03 mmol) was added to a solution of 250 mg (0.51 mmol) of $Re_2O_7$ in 5 mL of THF. The solution was stirred for 15 min and than cooled to -78 °C. 1.03 mmol of the amine was slowly added to the reaction mixture, which was than warmed to ambient temperature under continued stirring. Volatile materials were removed *in vacuo* and the residue extracted into pentane (3 x 10 mL), filtered and concentrated. By cooling to -22 °C, the products were obtained in crystalline form.

**(Di-isopropylamido)trioxorhenium (1).** The general procedure was applied using 0.14 mL (1.03 mmol) of diisopropylamine. The product was obtained as yellow crystals. Yield: 75 mg (23 %). - $^1$H-NMR (C$_6$D$_6$, 399.65 MHz): δ (ppm) = 3.40 (sept, *J*=6.4 Hz, 2H, (CH*$_3$*)$_2$-C*H*), 0.77 (d, *J*=6.4 Hz, 6H, C(C*H$_3$*)$_2$). - $^{13}$C{$^1$H}-NMR (C$_6$D$_6$, 100.40 MHz): δ (ppm) = 58.8 ((CH$_3$)$_2$-*C*H), 22.0 (C(*C*H$_3$)$_2$). - IR (KBr) ν (cm$^{-1}$) = 2973 s, 2839 m, 1460 m, 1396 m, 1374 m, 1261 m, 1140 m, 1099 s, 1022 m, 977 m (ν$_s$ [Re=O]), 934 s (ν$_{as}$ [Re=O]), 803 m. The data matches with ref. 15.

**(*N*-Cyclohexylisopropylamido)trioxorhenium (2).** The general procedure was applied using 0.17 mL (1.03 mmol) of *N*-cyclohexylisopropylamine. The product was obtained as yellow crystals. - Yield: 75 mg (19 %). - $^1$H-NMR (C$_6$D$_6$, 399.65 MHz): δ (ppm) = 3.45 (sept, *J*=6.5 Hz, 1H, (CH*$_3$*)$_2$-C*H*), 3.20 (tt, 1H), 1.41 (m, 4H), 1.19 (m, 3H), 0.88 (m, 1H), 0.82 (d, *J*=6.5 Hz, 6H, C(C*H$_3$*)$_2$), 0.74 (m, 2H). - $^{13}$C{$^1$H}-NMR (C$_6$D$_6$, 100.40 MHz): δ (ppm) = 67.3, 59.3 ((CH*$_3$*)$_2$-*C*H), 32.9, 25.6, 24.8, 22.2 (C(*C*H$_3$)$_2$). - IR (KBr) ν (cm$^{-1}$) = 2937 s, 2857 m, 1458 m, 1402 m, 1366 m, 1262 m, 1111 s, 1024 s, 976 m (ν$_s$ [Re=O]), 931 s (ν$_{as}$ [Re=O]), 802 s. - MS (80 eV, EI, 60 °C): *m/z* (%) = 375.2 (100) [M]$^+$, 360.2 (36) [M - Me]$^+$, 332.3 (43) [M – *i*Pr], 290.6 (51), 278.2 (66), 98.2 (107).

**(Dicyclohexylamido)trioxorhenium (3).** The general procedure was applied using 0.20 mL (1.03 mmol) of dicyclohexylamine. The product was obtained as orange crystals. - Yield: 70 mg (18 %). - $^1$H-NMR (C$_6$D$_6$, 399.65 MHz): δ (ppm) = 3.31 (tt, 2H, N-C*H*), 1.50 (m, 4H), 1.41 (m, 4H), 1.27 (m, 6H), 0.77 (m, 6H). - $^{13}$C{$^1$H}-NMR (C$_6$D$_6$, 100.40 MHz): δ (ppm) = 67.8 (N-*C*H), 33.1, 25.7, 24.8. - IR (KBr) ν (cm$^{-1}$) = 2939 s,

2853 m, 1447 m, 1399 m, 1341 m, 1254 m, 1153 m, 1080 m 1025 m, 951 s ($\nu_s$ [Re=O]), 921 s ($\nu_{as}$ [Re=O]), 849 m, 784 m. - MS (80 eV, EI, 70 °C): *m/z* (%) = 415.0 (100) [M]$^+$, 371.8 (6), 333.0 (28) [M – Cy], 290.1 (15), 138.3 (37), 98.0 (14), 81.3 (12), 55.1 (23), 40.8 (13).

**X-ray Crystallographic Studies of 1-3.**

Crystals suitable for X-ray crystallography were obtained from concentrated pentane solutions. A suitable crystal was covered in mineral oil (Aldrich) and mounted onto a glass fiber. The crystal was transferred directly to the -73 °C cold N$_2$ stream of a Stoe IPDS 2T or a Bruker CCD Apex 1000 diffractometer. Subsequent computations were carried out on an Intel Pentium IV PC.

All structures were solved by direct methods (SHELXS-97[33]). The remaining non-hydrogen atoms were located from successive difference Fourier map calculations. The refinements were carried out by using full-matrix least-squares techniques on *F*, minimizing the function $(F_o-F_c)^2$, where the weight is defined as $4F_0^2/2(F_o^2)$ and $F_o$ and $F_c$ are the observed and calculated structure factor amplitudes using the program SHELXL-97.[34] The hydrogen atom contributions were calculated, but not refined. The locations of the largest peaks in the final difference Fourier map calculation as well as the magnitude of the residual electron densities in each case were of no chemical significance. Crystallographic data (excluding structure factors) for the structures reported in this paper have been deposited with the Cambridge Crystallographic Data Centre as a supplementary publication no. CCDC-XXXX. Copies of the data can be obtained free of charge on application to CCDC, 12 Union Road, Cambridge CB21EZ, UK (fax: (+(44)1223-336-033; email: deposit@ccdc.cam.ac.uk).

**1**: $C_6H_{14}NO_3Re$ Monoclinic, $P2_1/n$ (No. 14); lattice constants $a$ = 7.8080(7), $b$ = 11.5114(12), $c$ = 11.2704(10) Å, $\beta$ = 102.380(7)°; $V$ = 989.4(2) Å$^3$, $Z$ = 4; $\mu$(Mo-$K_\alpha$) = 12.247 mm$^{-1}$; $\theta_{max.}$ = 25; 1746 [R$_{int}$ = 0.0685] independent reflections measured, of which 1582 were considered observed with $I > 2\sigma(I)$; max. residual electron density 1.955 and -3.504 e/A$^{-3}$; 105 parameters, $R1$ ($I > 2\sigma(I)$) = 0.0428; w$R2$ (all data) = 0.1163.

**2**: $C_9H_{18}NO_3Re$, Monoclinic, $Pbca$ (No. 61); lattice constants $a$ = 11.9102(6), $b$ = 17.5194(9), $c$ = 22.568(2) Å, $V$ = 4709.1(5) Å$^3$, $Z$ = 8; $\mu$(Mo-$K_\alpha$) = 10.306 mm$^{-1}$; $\theta_{max.}$ = 25.25; 4258 [R$_{int}$ = 0.0410] independent reflections measured, of which 3519 were considered observed with $I > 2\sigma(I)$; max. residual electron density 0.685 and -0.921 e/A$^{-3}$; 258 parameters, $R1$ ($I > 2\sigma(I)$) 0.0212; w$R2$ (all data) = 0.0393.

**3**: $C_{12}H_{22}NO_3Re$, Monoclinic, $Pna2_1$ (No. 33); lattice constants $a$ = 18.235(3), $b$ = 12.467(2), $c$ = 6.1486(11) Å, $V$ = 2604.0(3) Å$^3$, $Z$ = 4; $\mu$(Mo-$K_\alpha$) = 8.691 mm$^{-1}$; $\theta_{max.}$ = 30.5; 4243 [R$_{int}$ = 0.0211] independent reflections measured, of which 3828 were considered observed with $I > 2\sigma(I)$; max. residual electron density 0.952 and -0.577 e/A$^{-3}$; 256 parameters, $R1$ ($I > 2\sigma(I)$) = 0.0161; w$R2$ (all data) = 0.0344.

**Computational Details**

DFT calculations on the compounds **1**, **4**, **5** and **6** were carried out employing the BP86 exchange-correlation functional[35] and a TZ2P basis set as implemented in the ADF program package.[36] Relativistic effects were included by means of the scalar-relativistic ZORA Hamiltonian.[37] All structures were optimized without symmetry restrictions and the stationary points were verified to be true minima by analytical frequency calculations.

For the topological analysis of **4**, the electron density and its second derivative were calculated on a grid of points (0.01Å step) employing the program DGRID.[38]

In the case of **1** an additional single-point DFT calculation (BP86) on the final geometry of the all-electron calculations was carried out using effective core potentials containing the 1s-4s, 2p-4p, 3d-4d and the 4f subshells for the Re atom. The valence space consisting of the 5s-6s, 5p and 5d subshells was described by an uncontracted (5s, 5p, 4d) basis set.[39] For the C, O, N and H atoms a standard split-valence 6-31+G(d,p) basis set as implemented in GAUSSIAN03 was employed.[40] The electron density obtained from this calculation was analyzed [30] utilizing a locally modified version of AIMPAC.[41]

**Acknowledgment.** This work was supported by the Fonds der Chemischen Industrie the *Deutsche Forschungsgemeinschaft* (SPP1178) as well as NanoCat, an International Graduate Program within the Elitenetzwerk Bayern.

7.5 % were excluded. The full lists of results for both searches are given in the Supporting Information.

large core effective core potentials (ECPs) recovers the outermost shell and the presence of LICCs of complexes involving Period 6 elements. Hence, in our study small core ECP were employed to analyze the fine structure of the Laplacian. For further information see (a) G. Eickerling, R. Mastalerz, V. Herz, W. Scherer, H.-J. Himmel, M. Reiher *J. Chem. Theory Comput.* **2007**, *3*, 2182-2197. (b) G. Eickerling, M. Reiher *J. Chem. Theory Comput.* **2008**, *4*, 286-296 and (c) ref 20f.

[31] For a general discussion of the chemical nature of ligand-induced charge concentrations in $d^0$ type complexes see G. S. McGrady, A. Haaland, H. P. Verne, H. V. Volden, A. J. Downs, D. Shorokhov, G. Eickerling, W. Scherer *Chem. Eur. J.* **2005**, *11*, 4921-4934.

[32] (a) W. Scherer, P. Sirsch, M. Grosche, M. Spiegler, S. A. Mason, M. G. Gardiner *Chem. Commun*. **2001**, 2072-2073. (b) W. Scherer, P. Sirsch, D. Shorokhov, G. S. McGrady, S. A. Mason, M. Gardiner *Chem. Eur. J*. **2002**, *8*, 2324-2334. (c) L. Perrin, L. Maron, O. Eisenstein, M. F. Lappert *New. J. Chem.* **2003**, *27*, 121-127.

[33] G. M. Sheldrick, SHELXS-97, *Program of Crystal Structure Solution*, University of Göttingen, Germany, **1997**.

[34] G. M. Sheldrick, SHELXL-97, *Program of Crystal Structure Refinement*, University of Göttingen, Germany, **1997**.

[35] a) A. D. Becke *Phys. Rev. A* **1988**, *38*, 3098-3100; b) J. P. Perdew *Phys. Rev. B* **1986**, *33*, 8822-8824; c) J. P. Perdew *Phys. Rev. B* **1986**, *34*, 7406-7406.

[36] ADF2007.01, SCM, Theoretical Chemistry, Vrije Universiteit, Amsterdam, The Netherlands, http://www.scm.com; b) G. te Velde, F.M. Bickelhaupt, S.J.A. van Gisbergen, C. Fonseca Guerra, E.J. Baerends, J.G. Snijders, T. Ziegler *J. Comput. Chem.* **2001**, *22*, 931-967; c) C. Fonseca Guerra, J.G. Snijders, G. te Velde, E. J.

**Figure Caption:**

**Figure 1.** Solid-state structure of **1** showing the atom labeling scheme, omitting hydrogen atoms. Selected bond lengths [Å] or angles [°]: C1-N 1.492(10), C4-N 1.465(10), Re-N 1.893(6), Re-O1 1.694(8), Re-O2 1.708(6), Re-O3 1.710(6); O1-Re-O2 111.0(4), O1-Re-N1 108.2(3), O1-Re-O3 112.7(4), O2-Re-O3 106.5(4), O2-Re-N1 109.6(3), O3-Re-N1 108.7(3), C1-N-Re 125.6(5), C4-N-Re 114.7(5), C1-N-C4 119.6(6).

**Figure 2.** Solid-state structure of **2** showing the atom labeling scheme, omitting hydrogen atoms. Shown are both isomers located in the asymmetric unit. Selected bond lengths [Å] or angles [°]: C1-N1 1.498(5), C4-N1 1.482(5), Re1-N1 1.876(3), Re1-O1 1.696(3), Re1-O2 1.694(3), Re1-O3 1.704(4); O1-Re1-O2 109.9(2), O1-Re1-O3 108.0(2), O2-Re1-O3 111.7(2), O1-Re1-N1 110.3(2), O2-Re1-N1 109.1(2), O3-Re1-N1 107.8(2), C1-N1-Re1 113.4(3), C4-N1-Re1 127.7(3).
C10-N2 1.499(5), Re2-N2 1.876(3), Re2-O4 1.698(3), Re2-O5 1.699(3), Re2-O6 1.689(4); O4-Re2-O6 110.4(2), O5-Re2-O6 112.5(2), O4-Re2-O5 108.5(2), O6-Re2-N2 108.2(2), O4-Re2-N2 109.1(2), O5-Re2-N2 108.1(2), C10-N2-Re2 125.6(3), C13-N2-Re2 116.3(2).

**Figure 3.** Solid-state structure of **3** showing the atom labeling scheme, omitting hydrogen atoms. Selected bond lengths [Å] or angles [°]: C1-N 1.492(3), C7-N 1.493(3), Re-N 1.876(2), Re-O1 1.694(2), Re-O2 1.717(2), Re-O3 1.701(2); O1-Re-O2 110.4(2), O1-Re-O3 111.5(2), O2-Re-O3 106.8(1), O1-Re-N 110.3(1), O2-Re-N 108.8(1), O3-Re-N 109.0(1), C1-N-Re 114.0(2), C7-N-Re 127.7(2).

**Figure 4.** DFT models of **1** (left), **4** (middle) and **5** (right). Salient bond distances and angles are specified in [Å] and [°], respectively.

**Figure 5.** A search for the eclipsed structural fragments M-C-C-H and M-N-C-H (with M: $d^0$ transition metal) was performed employing the Cambridge Structural Database (CSD; ref 24) version 5.29 (November 2007). Solid triangles mark the entries of our theoretical models **1**, **4** and **5** while grey shaded triangles label the experimental models **1**, **2**, **3** and **4**.

**Figure 6.** (left) Contour map of the negative Laplacian of the electron density ($L(\mathbf{r}) = -\nabla^2\rho(\mathbf{r})$) of (a) **1** (left) in the ReNCH plane; the inset shows an envelope map of the negative Laplacian at the Re atom at 260 eÅ$^{-5}$. Contour levels in are drawn at 0, $\pm 2.0\times 10^n$, $\pm 4.0\times 10^n$, $\pm 8.0\times 10^n$ eÅ$^{-5}$, where $n = 0, \pm 3, \pm 2, \pm 1$; extra levels at 250 and 280 eÅ$^{-5}$ have been used. Positive and negative values are marked by solid and dashed lines, respectively. BCPs are marked by closed circles, while the bond path is shown by solid lines; $\rho(\mathbf{r})$ and $L(r)$ values at the CPs are specified in [eÅ$^{-3}$] and [eÅ$^{-5}$], respectively. The locations of the ligand-induced charge concentrations facing the C-H$_\beta$ bond are marked by arrows. (right) Envelope map of the negative Laplacian of the electron density of **4** at 200 eÅ$^{-5}$.

**Figure 7.** DFT models of **7** (left) and **8** (right). Salient bond distances and angles are specified in [Å] and [°], respectively.

**Figure 1.**

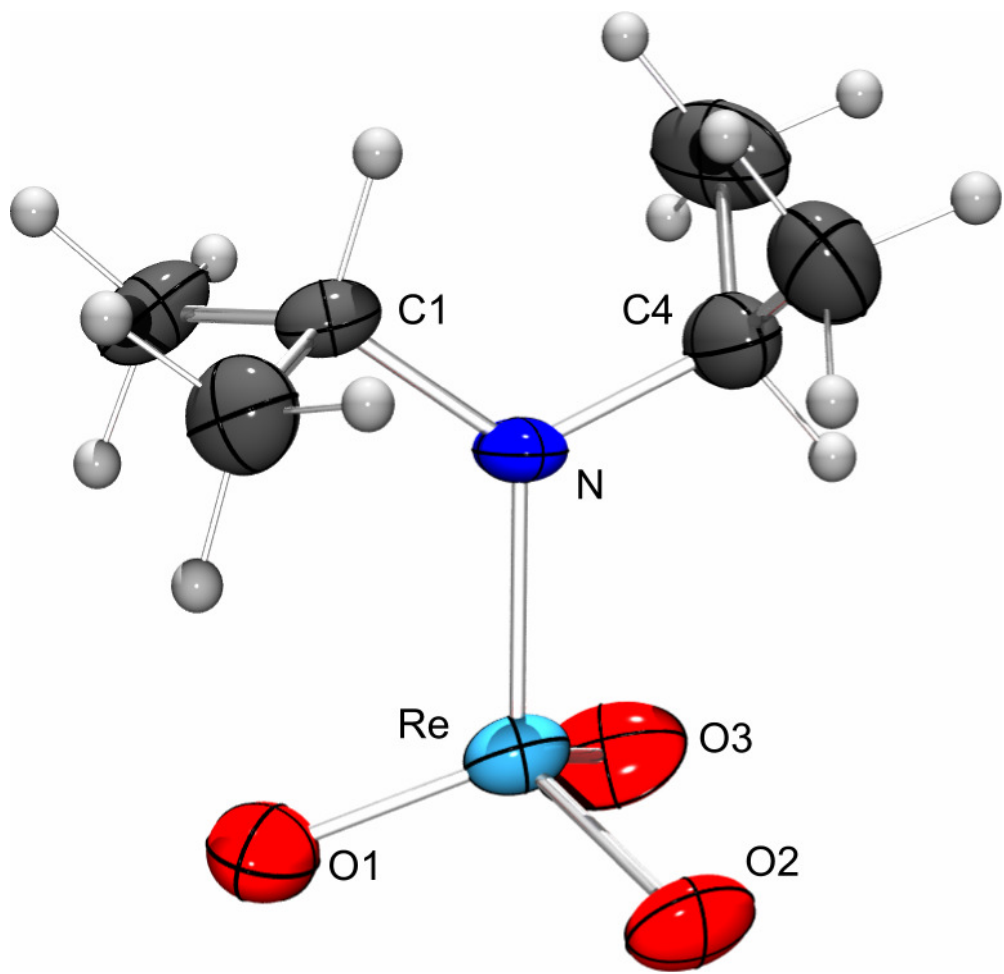

**Figure 2.**

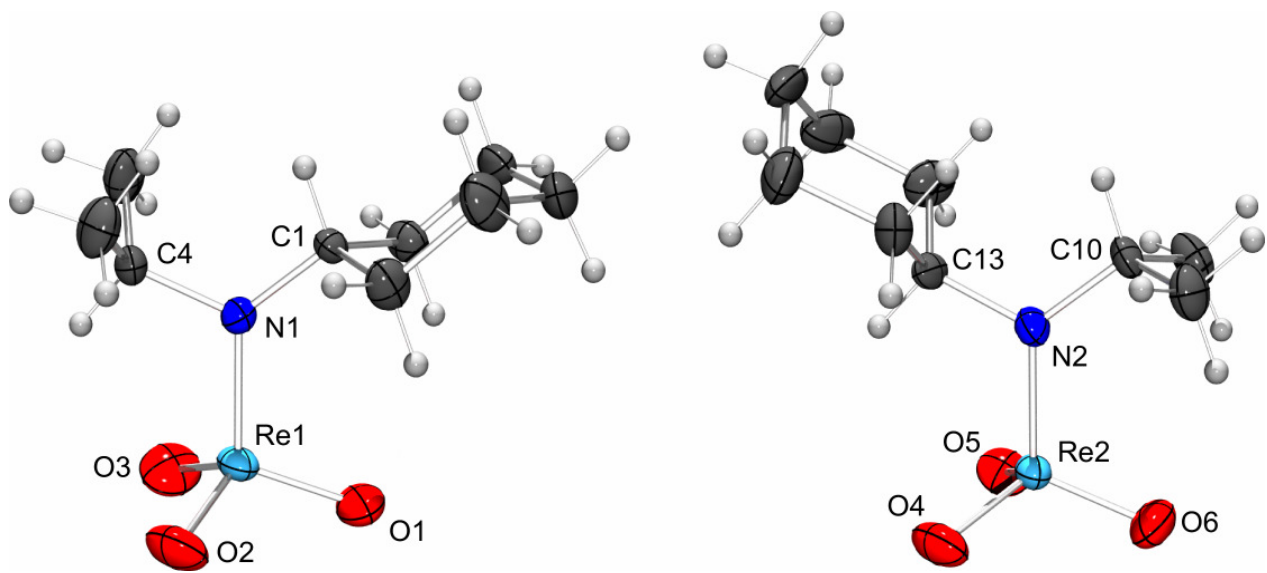

**Figure 3.**

**Figure 4.**

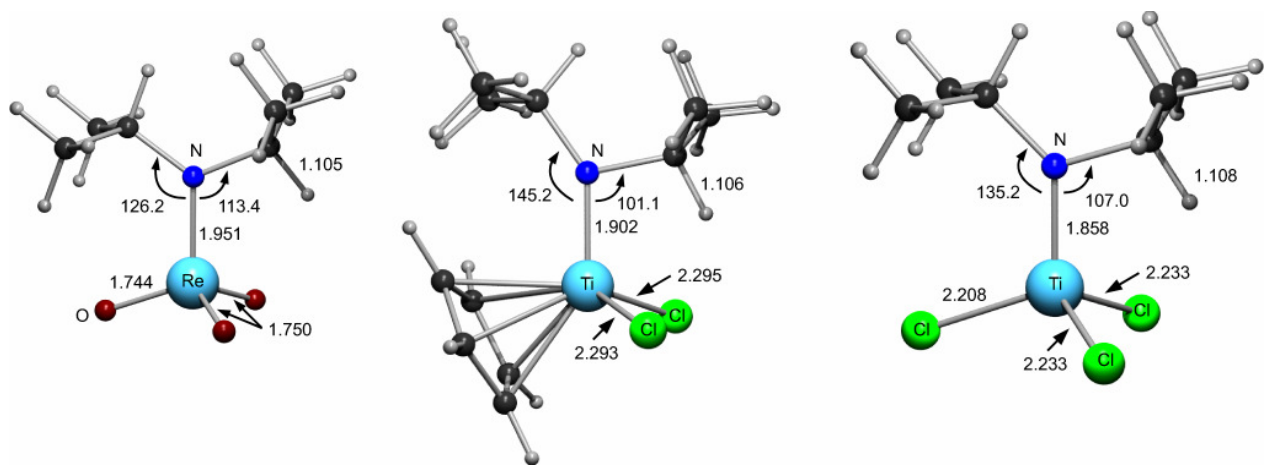

**Figure 5.**

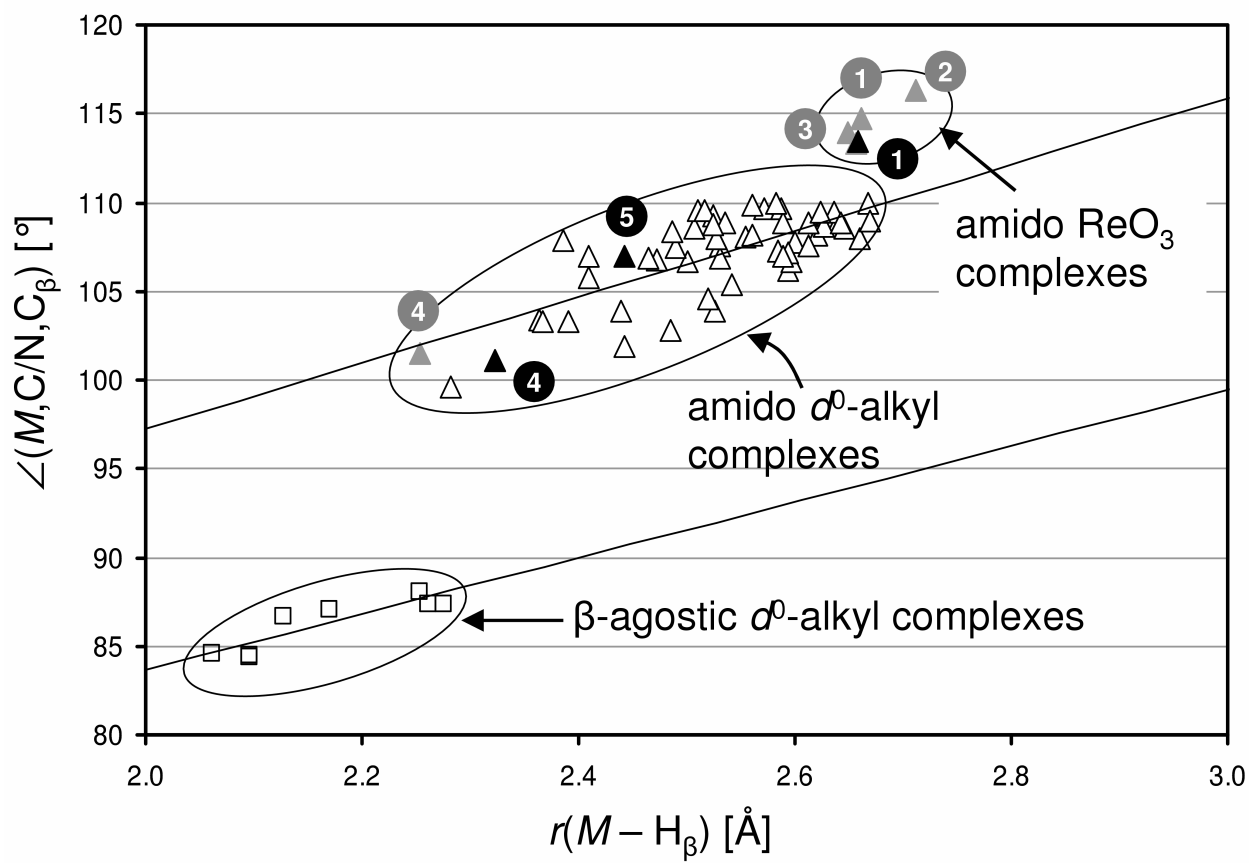

**Figure 6.**

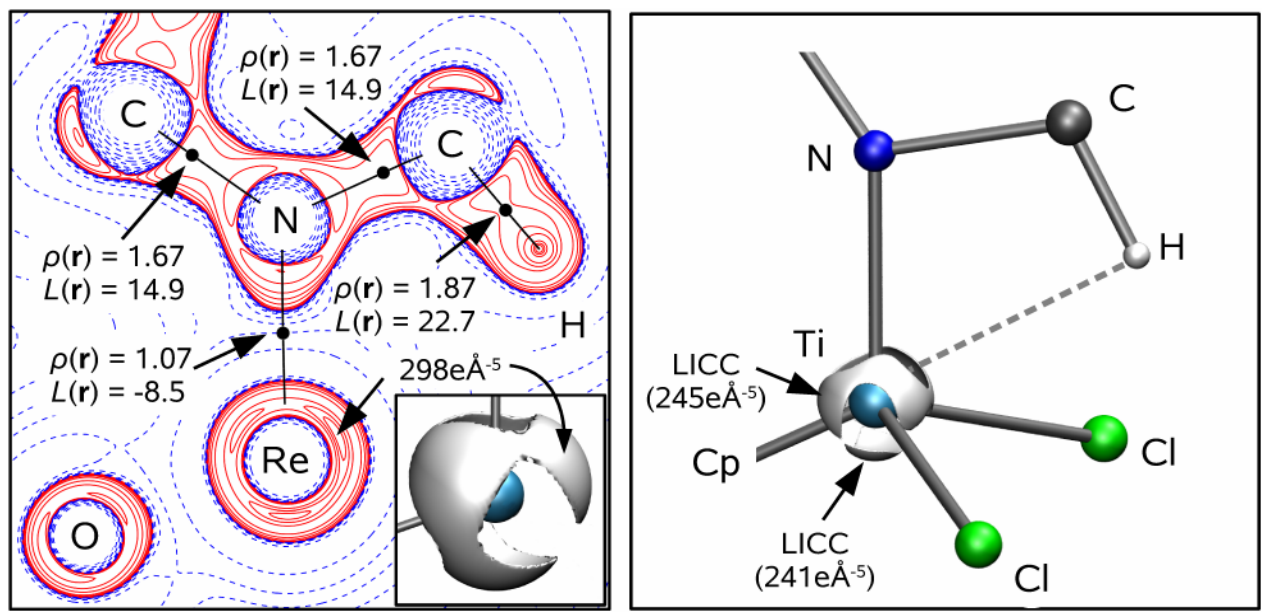

**Figure 7.**

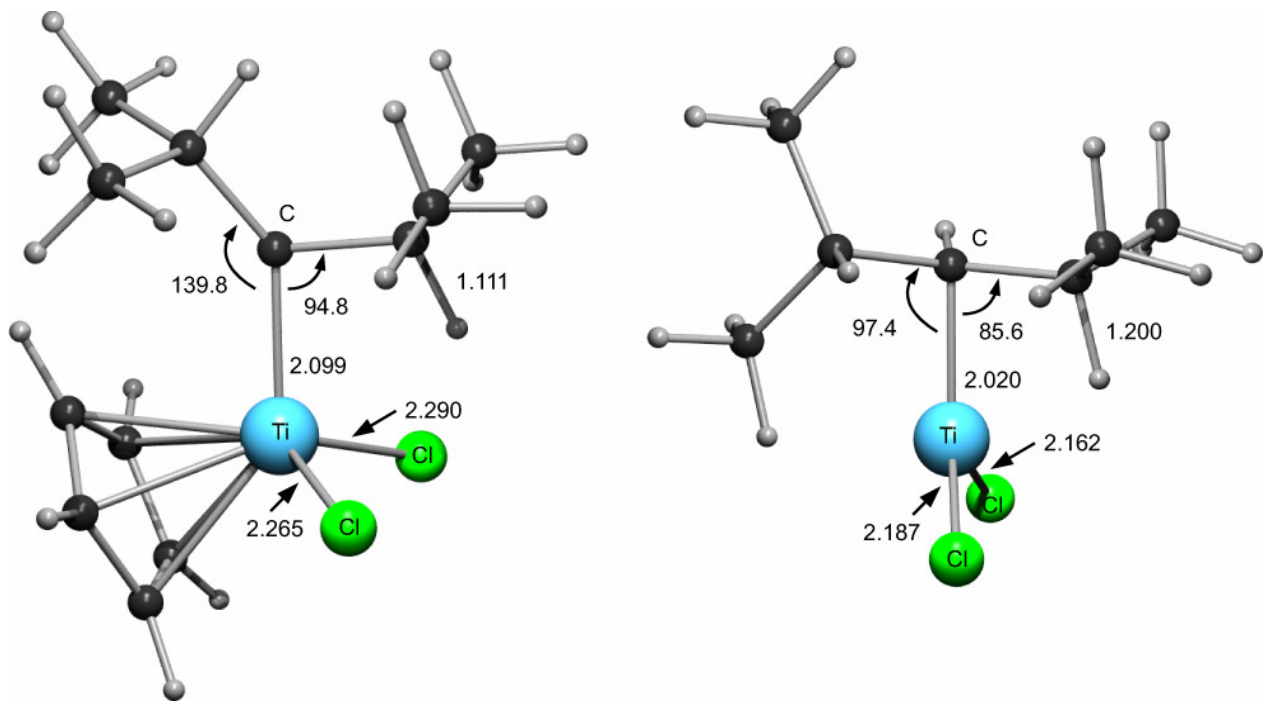

**Text for the table of contents:**

The amido rhenium trioxides of composition ($i$Pr$_2$N)ReO$_3$, ($i$PrCyN)ReO$_3$ and (Cy$_2$N)ReO$_3$ were synthesized and structurally characterized.

.

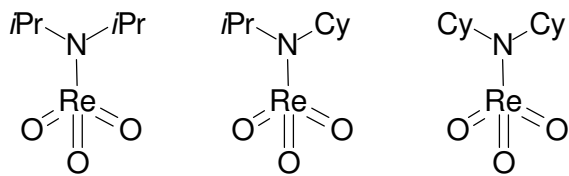